\documentclass[11pt, a4paper]{article}
\usepackage{moriond,epsfig}
\def\Dz{D^0}

\def\pep2{PEP-II}

\def\Rdcs{R_D}

\def\RdcsPM     {R_{\rm D}^{\pm}}

\def\xPrimeSq   {{x^{\prime}}^2}
\def\yPrimeSq   {{y^{\prime}}^2}
\def\xPrime     {x^{\prime}}

\def\yPrime     {y^{\prime}}
\def\yPrimeP    {y^{\prime+}}
\def\yPrimeM    {y^{\prime-}}
\def\xPrimePmSq {{x^{\prime}}^{2\pm}}

\def\yPrimePm   {y^{\prime\pm}}

\def\xPrimePSq  {{x^{\prime}}^{2+}}
\def\xPrimeMSq  {{x^{\prime}}^{2-}}
\def\xPrimePmSq  {{x^{\prime}}^{2\pm}}
\def\xPSQyP     {(\xPrimeSq,\,\yPrime)}

\def\AD         {A_{\rm D}}

\begin{document}
\vspace*{4cm}
\title{Evidence for $D^0$-$\kern 0.2em\overline{\kern -0.2em D}{}^0$ Mixing at Babar}

% Changed 040623 - All authors go here
\author{Kevin Flood}

\address{University of Wisconsin \\ Madison, WI, USA 53706}

\address{\bf{representing the Babar Collaboration}}

\maketitle\abstracts{We present evidence for $D^0$-$\kern 0.2em\overline{\kern -0.2em D}{}^0$ mixing in $D^0 \rightarrow K^{+}\pi^{-}$ decays from $384\mbox{\,fb}^{-1}$ of $e^+e^-$ colliding-beam data recorded near $\sqrt{s}=10.6\mathrm{\,Ge\kern -0.1em V}$ with the \mbox{\slshape B\kern-0.1em{\small A}\kern-0.1em B\kern-0.1em{\small A\kern-0.2em R}} detector at the PEP-II storage rings at SLAC. We find the mixing parameters ${x^{\prime}}^2 = [ -0.22 \pm 0.30 \hbox{ (stat.)}\pm 0.21 \hbox{ (syst.)}]\times 10^{-3}$ and $y^{\prime} = [9.7 \pm 4.4 \hbox{ (stat.)}\pm 3.1 \hbox{ (syst.)}] \times 10^{-3}$, and a correlation between them of $-0.94$. This result is inconsistent with the no-mixing hypothesis with a significance of 3.9 standard deviations. We measure $R_{\rm D}$, the ratio of doubly Cabibbo-suppressed to Cabibbo-favored decay rates, to be $[0.303\pm0.016\hbox{ (stat.)}\pm 0.010\hbox{ (syst.)}]\%$. We find no evidence for $C\!P$ violation.}

\section{Introduction}

The $\Dz$ and $\overline{\kern -0.2em D}{}^0$ mesons are flavor eigenstates which are invariant in strong interactions, but are subject to electroweak interactions that permit an initial flavor eigenstate to evolve into a time-dependent mixture of $\Dz$ and $\kern 0.2em\overline{\kern -0.2em D}{}^0$. In the Standard Model (SM), such oscillations proceed through both short-distance and non-perturbative long-distance amplitudes. The expected mixing rate mediated by down-type quark box diagrams~\cite{Datta:1985jx} and di-penguin~\cite{Petrov:1997ch} diagrams is ${\cal O}(10^{-8}-10^{-10})$, while the predicted range for non-perturbative long-distance contributions~\cite{Burdman:2003rs} is approximately bounded by the box diagram rate and the current experimental sensitivity of ${\cal O}(10^{-4})$. New physics predictions span the same large range.~\cite{Petrov:2003un} We present evidence for $D$~mixing consistent with these expectations and with previous experimental limits.~\cite{exp1}~\cite{exp2}~\cite{exp3}~\cite{exp4}~\cite{exp5}~\cite{exp6}~\cite{exp7} We also compare $\Dz$ and $\kern 0.2em\overline{\kern -0.2em D}{}^0$ samples separately, and find no evidence for $C\!P$ violation.

The mixing rate is characterized using the right-sign (RS), Cabibbo-favored (CF) decay~\footnote{The use of charge-conjugate modes is implied unless otherwise noted.} $D^0 \rightarrow K^-\pi^+$ and the wrong-sign (WS) decay $D^0 \rightarrow K^+\pi^-$. The WS final state can be produced either through a doubly Cabibbo-suppressed (DCS) tree-level decay or through mixing followed by a CF decay. The DCS decay has a small rate $\Rdcs$ of order $\tan^4\theta_{\rm C} \approx 0.3\%$ relative to CF decay, where $\theta_{\rm C}$ is the Cabibbo angle. We distinguish $\Dz$ and $\kern 0.2em\overline{\kern -0.2em D}{}^0$ by their production in the decay $D^{*+} \rightarrow \pi_s^+ D^0$. In RS decays, the $\pi_s^+$ and kaon have oppositely signed charges, while in WS decays the charge signs are the same. The time dependence of the WS decay rate is used to separate DCS from mixed decays.

Charm mixing is generally characterized by two dimensionless parameters, $x\equiv\Delta m/\Gamma$ and $y\equiv\Delta\Gamma/2\Gamma$, where $\Delta m=m_{2}-m_{1}$ ($\Delta\Gamma=\Gamma_{2}-\Gamma_{1}$) is the mass (width) difference between the two neutral $D$ mass eigenstates and $\Gamma$ is the average width. If either $x$ or $y$ is non-zero, then $D$ mixing will occur. We approximate the time dependence of the WS decay of a meson produced as a $\Dz$ at time~$t=0$ in the limit of small mixing ($|x|$, $|y| \ll 1$) and $C\!P$ conservation as~\cite{Blaylock:1995ay}
\begin{equation}
    \frac{T_{ws}}{e^{-\Gamma t}} \propto \Rdcs +
      \sqrt{\Rdcs}\yPrime\; \Gamma t +
      \frac{\xPrimeSq + \yPrimeSq}{4} (\Gamma t)^2\,,
\label{eq:Tws}
\end{equation}
where $\xPrime = x\cos\delta_{K\pi} + y \sin\delta_{K\pi}$, $\yPrime = -x\sin\delta_{K\pi} + y \cos\delta_{K\pi}$, and $\delta_{K\pi}$ is the strong phase between the DCS and CF amplitudes. We study both $C\!P$-conserving and $C\!P$-violating cases. For the $C\!P$-conserving case, we fit for the parameters $\Rdcs$, $\xPrimeSq$, and $\yPrime$. To search for $C\!P$ violation, we apply Eq.~\ref{eq:Tws} to the $\Dz$ and $\kern 0.2em\overline{\kern -0.2em D}{}^0$ samples separately, fitting for the parameters $\RdcsPM$, $\xPrimePmSq$, $\yPrimePm$ for $\Dz$($+$) and $\kern 0.2em\overline{\kern -0.2em D}{}^0$($-$) decays.

\section{Event Selection and Analysis}

We use 384 $\rm{fb}^{-1}$ of $e^+e^-$ colliding-beam data recorded near $\sqrt{s} = 10.6$ GeV with the Babar detector~\cite{Aubert:2001tu} at the PEP-II asymmetric-energy storage rings. We initially select signal candidates by combining oppositely-charged tracks identified as $K$ or $\pi$ using a likelihood-based particle identification algorithm, requiring the $K^{\pm}\pi^{\mp}$ invariant mass $1.81 < m_{K\pi} < 1.92$~GeV/$c^2$ and $e^+e^-$ center-of-mass frame (CM) momentum $p^*_{\rm{D0}} > 2.5$ GeV/$c$. We require the $\pi_s^+$ to have laboratory momentum $p > 0.1$ GeV/$c$ and CM momentum $p^* < 0.45$ GeV/$c$.

To obtain the proper decay time $t$ and its error $\sigma_t$ for each $\Dz$ candidate, we refit the $\Dz$ daughter tracks and the $\pi_s^+$, constraining the $\Dz$ daughters to originate from a common vertex while simultaneously requiring the $\Dz$ and $\pi_s^+$ to originate from a common vertex constrained by the position and size of the $e^+e^-$ interaction region. We require a refit $\chi^2$ probability $P(\chi^2) > 0.1\%$, $M(K\pi\pi_s^+)-m_{K\pi}$ mass difference $0.14 < \Delta m < 0.16$ GeV/$c^2$, proper decay time $-2 < t <4$ ps and proper decay time error $\sigma_t < 0.5$ ps. The nominal $\Dz$ mean proper lifetime is $\sim0.410$~ps~\cite{Yao:2006px} and the most probable value of $\sigma_t$ for signal events is 0.16~ps. If there are multiple signal candidates with overlapping tracks within an event, we retain only the candidate with the highest $P(\chi^2)$. After applying all selection criteria, we retain approximately 1,229,000~RS and 64,000~WS signal candidates. To avoid potential bias, the complete event selection and analysis procedures were finalized prior to examining the mixing results.

The mixing parameters are determined using an unbinned, extended maximum-likelihood fit to the RS and WS data samples over the four observables $m_{K\pi}$, $\Delta m$, $t$, and $\sigma_t$. The fit is performed in several stages. First, the shapes of the RS and WS signal and background probability density functions (PDFs) are determined from an initial 2-d fit to \{$m_{K\pi}, \Delta m$\}. These shapes are then fixed in subsequent fits. Next, the $D^0$ proper-time resolution function and lifetime are determined from a fit to the RS data using \{$m_{K\pi}, \Delta m$\} to separate the signal and background components. Finally, the WS data is analyzed using three different fit models. The first model assumes both $C\!P$ conservation and the absence of mixing, the second model allows mixing but no $C\!P$ violation, and the third model allows mixing and $C\!P$ violation.

The RS and WS \{$m_{K\pi}, \Delta m$\} distributions are described by four components: signal, random $\pi_s^+$, misreconstructed $\Dz$ and combinatorial background. The signal component has a characteristic peak in both $m_{K\pi}$ and $\Delta m$. The random $\pi_s^+$ component models reconstructed $\Dz$ decays combined with a random slow pion and has the same shape in $m_{K\pi}$ as signal events, but does not peak in $\Delta m$.  Misreconstructed $\Dz$ events have one or more of the $\Dz$ decay products either not reconstructed or reconstructed with the wrong particle hypothesis. They peak in $\Delta m$, but not in $m_{K\pi}$. For RS events, most of these are semileptonic $\Dz$\ decays. For WS events, the main contribution is RS $\Dz \rightarrow K^- \pi^+$ decays where the $K^-$ and the $\pi^+$ are misidentified as $\pi^-$ and $K^+$, respectively. Combinatorial background events comprise the remainder of events and do not exhibit any peaking structure in $m_{K\pi}$ or $\Delta m$.

\begin{figure}[phtb]
  \centering
  \centerline{%
    \includegraphics[width=0.5\linewidth, clip=]{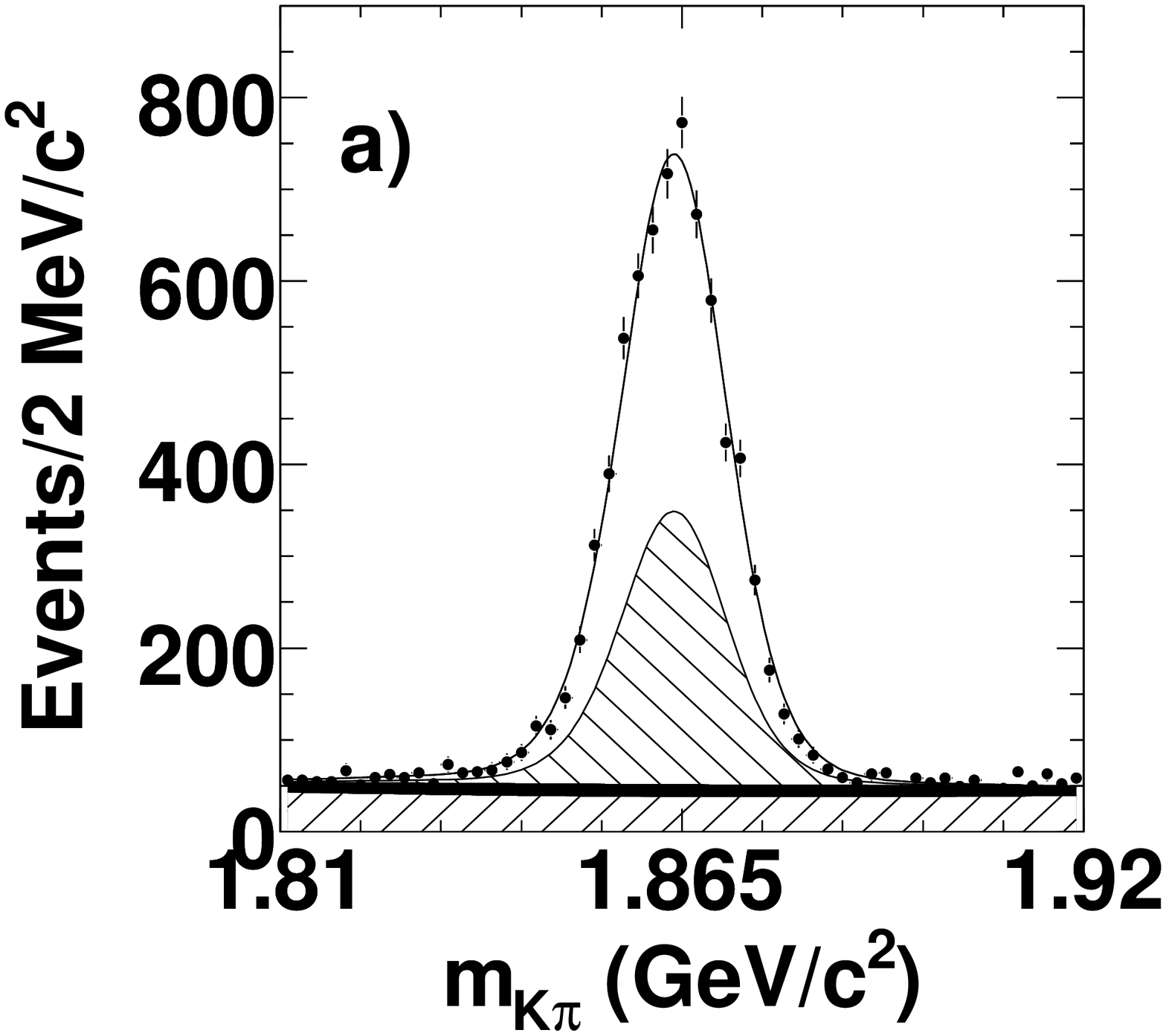}
    \includegraphics[width=0.5\linewidth, clip=]{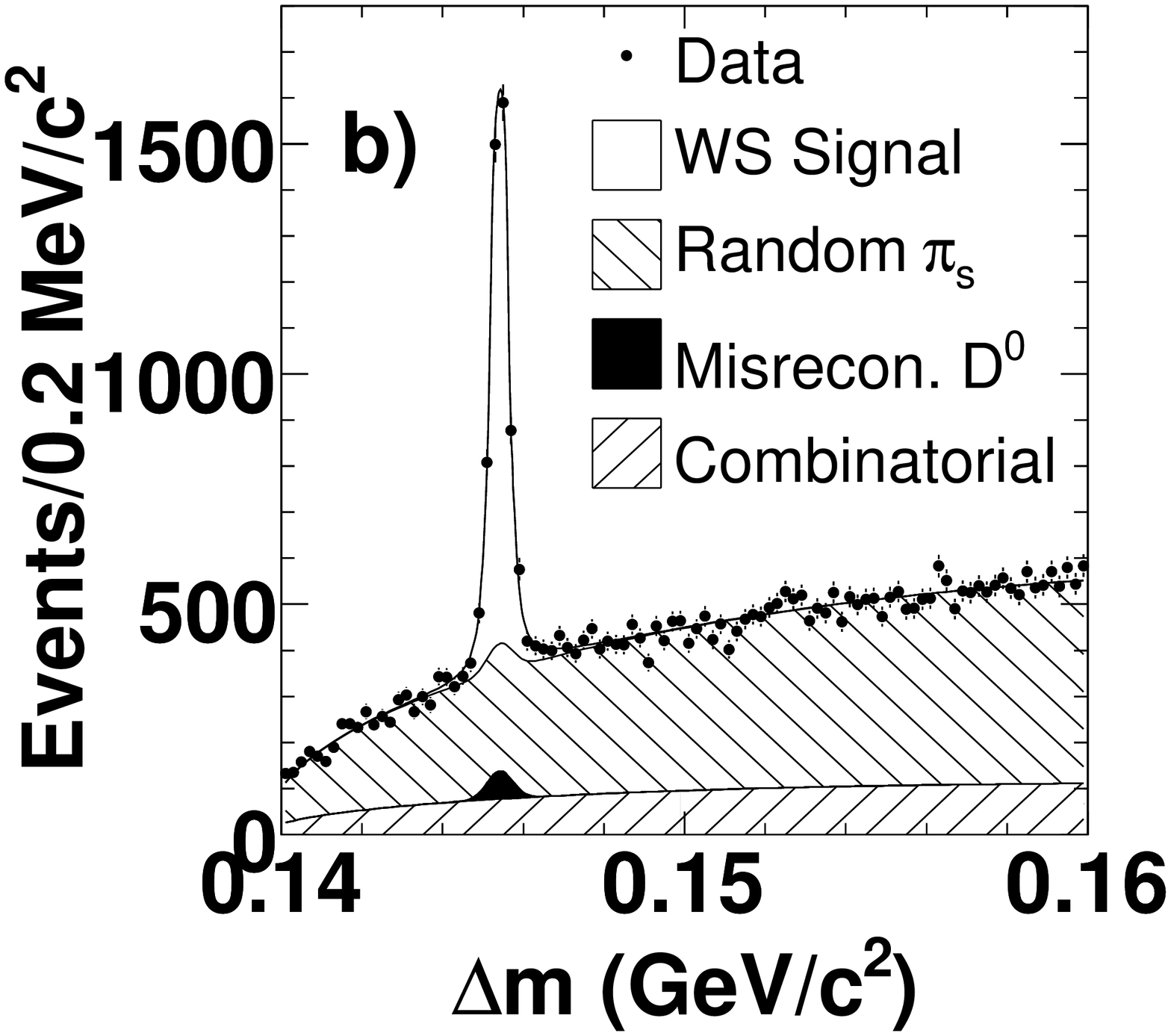}
  }
  \caption{a) $m_{K\pi}$ for WS candidates with $0.1445< \Delta m <0.1465$ GeV/$c^2$, and b) $\Delta m$ for WS candidates with $1.843 < m_{K\pi} < 1.883$  GeV/$c^2$. The fitted PDFs are overlaid.}
  \label{fig:r18Data_MdMfit}
%  \smallskip
\end{figure}

We fit the RS and WS data samples simultaneously to determine the PDF parameters describing the signal and random $\pi_s^+$ event class shapes shared between RS and WS datasets. We find $1,141,500\pm 1,200$ RS signal events and $4,030\pm 90$ WS signal events. The dominant background component is the random $\pi_s^+$ background. Projections of the WS data and fit are shown in Fig.~\ref{fig:r18Data_MdMfit}.

The measured proper-time distribution for the RS signal is described by an exponential function convolved with a resolution function whose parameters are determined by the fit to the data. The resolution function is the sum of three Gaussians with widths proportional to the estimated event-by-event proper-time uncertainty $\sigma_t$. The random $\pi_s^+$ background is described by the same proper-time distribution as signal events, since the slow pion has little weight in the vertex fit. The proper-time distribution of the combinatorial background is described by a sum of two Gaussians, one of which has a power-law tail to account for a small long-lived component.  The combinatorial background and real $\Dz$ decays have different $\sigma_t$ distributions, as determined from data using a background-subtraction technique~\cite{Pivk:2004ty} based on the \{$m_{K\pi}, \Delta m$\} fit.

The fit to the RS proper-time distribution is performed over all events in the full $m_{K\pi}$ and $\Delta m$ region. The PDFs for signal and background in $m_{K\pi}$ and $\Delta m$ are used in the proper-time fit with all parameters fixed to their previously determined values.  The fitted $\Dz$ lifetime is found to be consistent with the world-average lifetime.~\cite{Yao:2006px}

\begin{figure}[phtb]
\centering
\centerline{%
  \includegraphics[width=0.5\linewidth, clip=]{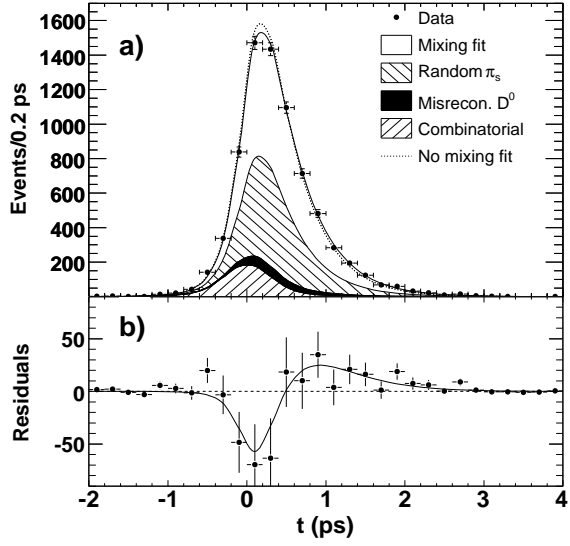}
}
\caption{a) Projections of the proper-time distribution of combined $\Dz$ and $\kern 0.2em\overline{\kern -0.2em D}{}^0$ WS candidates and fit result integrated over the signal region $1.843 < m_{K\pi} < 1.883$ GeV/$c^2$ and $0.1445 < \Delta m <0.1465$ GeV/$c^2$. The result of the fit allowing (not allowing) mixing but not $C\!P$ violation is overlaid as a solid (dashed) curve. b) The points represent the difference between the data and the no-mixing fit. The solid curve shows the difference between fits with and without mixing.}
\label{fig:histTimeBiasWSR18Data}
\end{figure}

\begin{figure}[phtb]
  \centering
  \centerline{%
    \includegraphics[width=0.5\linewidth, clip=]{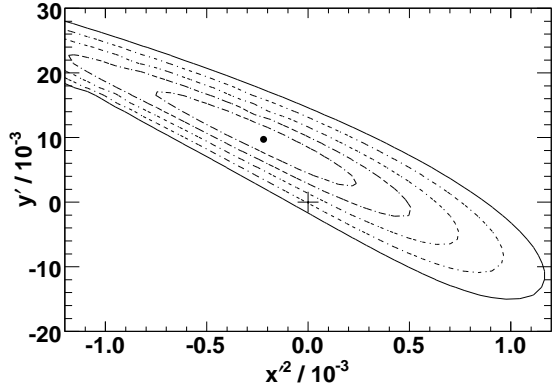}
}
\caption{The central value (point) and confidence-level (CL) contours for $1-\mbox{CL}=0.317\ (1\sigma)$, $4.55\times10^{-2}\ (2\sigma)$, $2.70\times10^{-3}\ (3\sigma)$, $6.33\times10^{-5}\ (4\sigma)$ and $5.73\times10^{-7}\ (5\sigma)$, calculated from the change in the value of $-2\ln{\cal L}$ compared with its value at the minimum. Systematic uncertainties are included. The no-mixing point is shown as a plus sign~($+$).}
\label{fig:CPContour}
\end{figure}

The measured proper-time distribution for the WS signal is modeled by Eq.~\ref{eq:Tws} convolved with the resolution function determined in the RS proper-time fit. The random $\pi_s^+$ and misreconstructed $\Dz$ backgrounds are described by the RS signal proper-time distribution since they are real $\Dz$ decays. The proper-time distribution for WS data is shown in Fig.~\ref{fig:histTimeBiasWSR18Data}. The fit results with and without mixing are shown as the overlaid curves. The fit with mixing provides a substantially better description of the data than the fit with no mixing. The significance of the mixing signal is evaluated based on the change in negative log likelihood with respect to the minimum. Figure~\ref{fig:CPContour} shows confidence-level (CL) contours calculated from the change in log likelihood ($-2\Delta\ln{\cal L}$) in two dimensions ($\xPrimeSq$ and $\yPrime$) with systematic uncertainties included.  The likelihood maximum is at the unphysical value of $\xPrimeSq=-2.2\times10^{-4}$ and $\yPrime = 9.7 \times 10^{-3}$. The value of $-2\Delta\ln{\cal L}$ at the most likely point in the physically allowed region ($\xPrimeSq=0$ and $\yPrime=6.4 \times 10^{-3}$) is $0.7$~units.  The value of $-2\Delta\ln{\cal L}$ for no-mixing is $23.9$~units. Including the systematic uncertainties, this corresponds to a significance equivalent to 3.9~standard deviations ($1-\mbox{CL}=1\times10^{-4}$) and thus constitutes evidence for mixing. The fitted values of the mixing parameters and $\Rdcs$, along with errors, are listed in Table~\ref{tab:results}.  The correlation coefficient between the $\xPrimeSq$ and $\yPrime$ parameters is $-0.94$.

Allowing for the possibility of $C\!P$ violation, we calculate the values of $\Rdcs = \sqrt{\Rdcs^+\Rdcs^-}$ and $\AD = (\Rdcs^{+} - \Rdcs^{-})/(\Rdcs^{+} + \Rdcs^{-})$ listed in Table~\ref{tab:results}, from the fitted $\Rdcs^{\pm}$ values.  The best fit points $(\xPrimePmSq,\yPrimePm)$ shown in Table~\ref{tab:results} are more than three standard deviations away from the no-mixing hypothesis. The shapes of the $(\xPrimePmSq,\yPrimePm)$ CL contours are similar to those shown in Fig.~\ref{fig:CPContour}. All cross-checks indicate that the close agreement between the separate $\Dz$ and $\kern 0.2em\overline{\kern -0.2em D}{}^0$ fit results is coincidental.

\begin{figure}[phtb]
  \centering
  \centerline{%
    \includegraphics[width=0.5\linewidth, clip=]{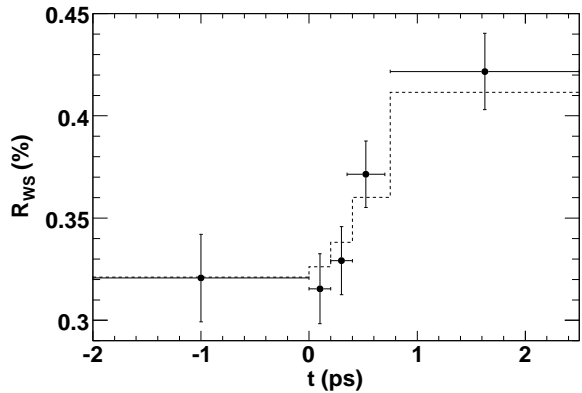}
  }
\caption{The WS branching fractions from independent ${m_{K\pi}, \Delta m}$ fits to slices in measured proper time (points). The dashed line shows the expected wrong-sign rate as determined from the mixing fit shown in Fig.~\ref{fig:histTimeBiasWSR18Data}. The $\chi^2$ with respect to expectation from the mixing fit is 1.5; for the no-mixing hypothesis (a constant WS rate), the $\chi^2$ is 24.0.}
\label{fig:RwsTimeBins}
\end{figure}

As a cross-check of the mixing signal, we perform independent \{$m_{K\pi}, \Delta m$\{ fits with no shared parameters for intervals in proper time selected to have approximately equal numbers of RS candidates. Figure~\ref{fig:RwsTimeBins} shows the resulting fitted WS branching fractions growing with increasing proper time. The slope of a linear fit to the data points is consistent with the measured mixing parameters and inconsistent with the no-mixing hypothesis.

\begin{table}[t]
\caption{Results from the different fits. The first uncertainty listed is statistical and the second systematic.}
\label{tab:results}
\vspace{0.4cm}
\begin{center}
\begin{tabular}{|l|c|c|}
\hline
Fit type & Parameter & Fit Results ($/10^{-3}$) \\ \hline
No $C\!P$ violation or mixing & $\Rdcs$       & $3.53 \pm 0.08 \pm 0.04$ \\ \hline
No $C\!P$ violation &           $\Rdcs$       & $3.03 \pm 0.16 \pm 0.10$ \\
                &           $\xPrimeSq$   & $-0.22 \pm 0.30 \pm 0.21$ \\
                &           $\yPrime$     & $9.7 \pm 4.4 \pm 3.1$ \\ \hline
$C\!P$ violation allowed      & $\Rdcs$       & $3.03 \pm 0.16 \pm 0.10$  \\
    & $\AD$        & $-21    \pm 52     \pm 15$  \\
    & $\xPrimePSq$ & $-0.24  \pm  0.43  \pm  0.30 $ \\
    & $\yPrimeP$   & $ 9.8   \pm  6.4   \pm  4.5  $ \\
    & $\xPrimeMSq$ & $-0.20  \pm  0.41  \pm  0.29 $ \\
    & $\yPrimeM$   & $ 9.6   \pm  6.1   \pm  4.3  $ \\ \hline
\end{tabular}
\end{center}
\end{table}

We validated the fitting procedure on simulated data samples using both MC samples with the full detector simulation and large parameterized MC samples. In all cases we found the fit to be unbiased. As a further cross-check, we performed a fit to the RS data proper-time distribution allowing for mixing in the signal component; the fitted values of the mixing parameters are consistent with no mixing. In addition we found the staged fitting approach to give the same solution and confidence regions as a simultaneous fit in which all parameters are allowed to vary.

In evaluating systematic uncertainties in $\Rdcs$ and the mixing parameters we considered variations in the fit model and in the selection criteria. We also considered alternative forms of the $m_{K\pi}$, $\delta m$, proper time, and $\sigma_t$ PDFs.  We varied the $t$ and $\sigma_t$ requirements. In addition, we considered variations that keep or reject all $D^{*+}$ candidates sharing tracks with other candidates. For each source of systematic error, we compute the significance $s_i^2=2\left[\ln{\cal L}\xPSQyP-\ln{\cal L}(\xPrimeSq_i, \yPrime_i)\right]/2.3$, where $\xPSQyP$ are the parameters obtained from the standard fit, $(\xPrimeSq_i, \yPrime_i)$ the parameters from the fit including the $i^{th}$ systematic variation, and $\cal L$ the likelihood of the standard fit. The factor 2.3 is the 68\% confidence level for 2 degrees of freedom. To estimate the significance of our results in $\xPSQyP$, we reduce $-2\Delta\ln{\cal L}$ by a factor of $1+\Sigma s_i^2=1.3$ to account for systematic errors. The largest contribution to this factor, $0.06$, is due to uncertainty in modeling the long decay time component from other $D$ decays in the signal region. The second largest component, $0.05$, is due to the presence of a non-zero mean in the proper time signal resolution PDF. The mean value is determined in the RS proper time fit to be 3.6~fs and is due to small misalignments in the detector. The error of $15\times 10^{-3}$ on $\AD$ is primarily due to uncertainties in modeling the differences between $K^+$ and $K^-$ absorption in the detector.

We have presented evidence for $\Dz$-$\kern 0.2em\overline{\kern -0.2em D}{}^0$ mixing. Our result is inconsistent with the no-mixing hypothesis at a significance of 3.9 standard deviations. We measure $\yPrime=[9.7 \pm 4.4 \hbox{ (stat.)}\pm 3.1 \hbox{ (syst.)}] \times 10^{-3}$, while $\xPrimeSq$ is consistent with zero. We find no evidence for $C\!P$ violation and measure $\Rdcs$ to be $[0.303\pm0.016\hbox{ (stat.)}\pm 0.010\hbox{ (syst.)}]\%$.  The result is consistent with SM estimates for mixing.

\section*{Acknowledgments}

We are grateful for the extraordinary contributions of our \pep2\ colleagues in achieving the excellent luminosity and machine conditions that have made this work possible. The success of this project also relies critically on the expertise and dedication of the computing organizations that support Babar. The collaborating institutions wish to thank SLAC for its support and the kind hospitality extended to them. This work is supported by the US Department of Energy and National Science Foundation, the Natural Sciences and Engineering Research Council (Canada), the Commissariat \`a l'Energie Atomique and Institut National de Physique Nucl\'eaire et de Physique des Particules (France), the Bundesministerium f\"ur Bildung und Forschung and Deutsche Forschungsgemeinschaft (Germany), the Istituto Nazionale di Fisica Nucleare (Italy), the Foundation for Fundamental Research on Matter (The Netherlands), the Research Council of Norway, the Ministry of Science and Technology of the Russian Federation, Ministerio de Educaci\'on y Ciencia (Spain), and the Science and Technology Facilities Council (United Kingdom). Individuals have received support from the Marie-Curie IEF program (European Union) and the A. P. Sloan Foundation.

\end{document}